# Prime Focus Spectrograph for the Subaru telescope: massively multiplexed optical and near-infrared fiber spectrograph

Hajime Sugai*[a], Naoyuki Tamura[a], Hiroshi Karoji[a], Atsushi Shimono[a], Naruhisa Takato[b], Masahiko Kimura[c], Youichi Ohyama[c], Akitoshi Ueda[d], Hrand Aghazarian[e], Marcio Vital de Arruda[f], Robert H. Barkhouser[g], Charles L. Bennett[g], Steve Bickerton[a], Alexandre Bozier[h], David F. Braun[e], Khanh Bui[i], Christopher M. Capocasale[e], Michael A. Carr[j], Bruno Castilho[f], Yin-Chang Chang[c], Hsin-Yo Chen[c], Richard C.Y. Chou[c], Olivia R. Dawson[e], Richard G. Dekany[k], Eric M. Ek[e], Richard S. Ellis[i], Robin J. English[e], Didier Ferrand[h], Décio Ferreira[f], Charles D. Fisher[e], Mirek Golebiowski[g], James E. Gunn[j], Murdock Hart[g], Timothy M. Heckman[g], Paul T. P. Ho[c], Stephen Hope[g], Larry E. Hovland[e], Shu-Fu Hsu[c], Yen-Shan Hu[c], Pin Jie Huang[c], Marc Jaquet[h], Jennifer E. Karr[c], Jason G. Kempenaar[e], Matthew E. King[e], Olivier Le Fèvre[h], David Le Mignant[h], Hung-Hsu Ling[c], Craig Loomis[j], Robert H. Lupton[j], Fabrice Madec[h], Peter Mao[i], Lucas Souza Marrara[f], Brice Ménard[g], Chaz Morantz[e], Hitoshi Murayama[a], Graham J. Murray[l], Antonio Cesar de Oliveira[f], Claudia Mendes de Oliveira[m], Ligia Souza de Oliveira[f], Joe D. Orndorff[g], Rodrigo de Paiva Vilaça[f], Eamon J. Partos[e], Sandrine Pascal[h], Thomas Pegot-Ogier[h], Daniel J. Reiley[i], Reed Riddle[i], Leandro Santos[f], Jesulino Bispo dos Santos[f], Mark A. Schwochert[e], Michael D. Seiffert[e,i], Stephen A. Smee[g], Roger M. Smith[k], Ronald E. Steinkraus[e], Laerte Sodré Jr[m], David N. Spergel[j], Christian Surace[h], Laurence Tresse[h], Clément Vidal[h], Sebastien Vives[h], Shiang-Yu Wang[c], Chih-Yi Wen[c], Amy C. Wu[e], Rosie Wyse[g], Chi-Hung Yan[c]

[a]Kavli Institute for the Physics and Mathematics of the Universe (WPI), The University of Tokyo, 5-1-5, Kashiwanoha, Kashiwa, 277-8583, Japan;
[b]Subaru Telescope, National Astronomical Observatory of Japan, 650 North A`ohoku Pl., Hilo, Hawaii, 96720, USA;
[c]Institute of Astronomy and Astrophysics, Academia Sinica, P.O. Box 23-141, Taipei, Taiwan;
[d]National Astronomical Observatory of Japan, 2-21-1 Osawa, Mitaka, Tokyo 181-8588, Japan;
[e]Jet Propulsion Laboratory, 4800 Oak Grove Drive, Pasadena, CA 91109, USA;
[f]Laboratório Nacional de Astrofisica, MCTI, Rua Estados Unidos, 154, Bairro das Nações, Itajubá, MG, Brazil;
[g]Department of Physics and Astronomy, Johns Hopkins University, 3400 North Charles Street, Baltimore, MD 21218, USA;
[h]Aix Marseille Université, CNRS, LAM (Laboratoire d'Astrophysique de Marseille) UMR 7326, 13388, Marseille, France;
[i]Astronomy Department, California Institute of Technology, 1200 East California Blvd, Pasadena, CA 91125, USA;
[j]Department of Astrophysical Sciences, Princeton University, Princeton, NJ, 08544, USA;
[k]Caltech Optical Observatories, 1201 East California Blvd., Pasadena, CA 91125 USA;
[l]Centre For Advanced Instrumentation, Durham University, Physics Dept, Rochester Bldg, South Road, Durham DH1 3LE, UK;
[m]Instituto de Astronomia, Geofisica e Ciencias Atmosfericas, Universidade de São Paulo, Rua do Matão, 1226 - Cidade Universitária - 05508-090, Brazil

* hajime.sugai@ipmu.jp; phone 81 4 7136-6551; fax 81 4 7136-6576; http://member.ipmu.jp/hajime.sugai/

## ABSTRACT

The Prime Focus Spectrograph (PFS) is an optical/near-infrared multifiber spectrograph with 2394 science fibers distributed across a 1.3-deg diameter field of view at the Subaru 8.2-m telescope. The wide wavelength coverage from 0.38 μm to 1.26 μm, with a resolving power of 3000, simultaneously strengthens its ability to target three main survey programs: cosmology, galactic archaeology and galaxy/AGN evolution. A medium resolution mode with a resolving power of 5000 for 0.71 μm to 0.89 μm will also be available by simply exchanging dispersers. We highlight some of the technological aspects of the design. To transform the telescope focal ratio, a broad-band coated microlens is glued to each fiber tip. A higher transmission fiber is selected for the longest part of the cable system, optimizing overall throughput; a fiber with low focal ratio degradation is selected for the fiber-positioner and fiber-slit components, minimizing the effects of fiber movements and fiber bending. Fiber positioning will be performed by a positioner consisting of two stages of piezo-electric rotary motors. The positions of these motors are measured by taking an image of artificially back-illuminated fibers with the metrology camera located in the Cassegrain container; the fibers are placed in the proper location by iteratively measuring and then adjusting the positions of the motors. Target light reaches one of the four identical fast-Schmidt spectrograph modules, each with three arms. The PFS project has passed several project-wide design reviews and is now in the construction phase.

## 1. INTRODUCTION

The Prime Focus Spectrograph (PFS)[1] is an optical/near-infrared multifiber spectrograph with 2394 science fibers (Fig. 1), and is scheduled to be mounted on the Subaru 8.2-m telescope. The position of each fiber is controlled by a fiber positioner, called "Cobra," consisting of two-staged piezo-electric rotary squiggle motors. Each fiber positioner unit samples a circular patrol region of 9.5-mm diameter at the telescope focal plane, and these positioners are arrayed into a hexagonal pattern with 8-mm central distances between adjacent positioners. The whole array extends to a single hexagonal field of view (FOV) whose effective diameter is 1.3 deg. The hexagonal FOV has an advantage for efficient tiling of survey areas, compared with e.g., a circular FOV. The fibers are divided into four groups, each of which enters one of the four spectrograph modules in a slit comprising 600 or 597 science fibers in a single row. Each spectrograph module has three color arms to cover a wide wavelength region from 0.38 μm to 1.26 μm, with a mean resolving power $\lambda/\delta\lambda$ of 3000. This resolution is optimal for spectroscopic surveys of fainter galaxies and stars, targeting cosmology, Galactic archaeology and galaxy/AGN evolution. To simplify the design of the spectrograph, the F/2.2 beam from the telescope wide field corrector (WFC) is increased to F/2.8 by a microlens bonded to each fiber. A double Schmidt design is used for the spectrograph modules, which are F/2.5 at the fiber side and F/1.09 at the detector side. These apertures are slightly larger than the nominal F/#, allowing for some focal ratio degradation (FRD) in the fiber. A medium spectral resolution mode with resolving power of 5000 for the red arms (0.71 μm to 0.89 μm) is also included. This addition particularly benefits important Galactic archaeology studies. This new mode is realized by using a simple grating/grism exchange mechanism without changing any other design elements. The basic characteristics of PFS are summarized in Table 1. The total efficiency from the atmosphere to the detector is expected to be around ~10%–20%,[1] depending on the wavelength and observing conditions such as telescope elevation and fiber position in FOV.

Together, PFS and hyper suprime-cam (HSC)[2] comprise the Subaru measurement of the images and redshifts project. PFS will accomplish the spectroscopic aspect of this project, and HSC has been implementing the imaging part since first light in 2012. The scientific case for the PFS surveys is described in an earlier paper.[3] PFS has now entered the construction phase, and Fig. 2 shows the integration and test flow. Generally speaking, PFS comprises two main components: the prime focus instrument (PFI) and the spectrographs, with the fiber system connecting the two parts. This paper provides the first technical overview of the instrument, describing the essential components as well as their intensive test results. Further details on individual PFS components can be found in separate, componentspecific SPIE conference papers viz.: fiber system,[4–9] fiber positioner[10] and prime focus instrument,[11] spectrograph,[12–14] Dewar and detector,[15–18] and metrology camera.[19] The subsystems are discussed in roughly the same order as the science light path through the instrument. The Subaru telescope and WFC focus light through a microlens; this microlens, which is described in Sec. 2.1, increases the F/# of the beam to ease the design requirements on the spectrographs. The light then enters the multimode fibers, which are part of a fiber system described in Sec. 2.2 that also includes connectors and fiber arrays. In the telescope focal plane, these fibers are mounted in a series of positioners, described in Sec. 2.3, which connect the fibers to the celestial images. The fibers transmit the science light to one of the four spectrograph modules, described in Sec. 2.4. The spectrograph modules also provide a means for back-illumination, which enables the metrology camera system, described in Sec. 2.5, to measure the positions of the fibers. The relevant software and observational execution processes are described in Sec. 2.6. PFS technical first light is scheduled on Subaru in early 2018 with the start of the associated Subaru strategic program survey expected in 2019.

The PFS collaboration is organized from Kavli IPMU (WPI) with PI Hitoshi Murayama and a PFS project office. Other members of the consortium include astronomers and engineers from Caltech/JPL, Princeton and JHU in the USA, LAM in France, USP/LNA in Brazil, and ASIAA in Taiwan. In 2014 January, NAOJ/Subaru formally joined the project, and MPA, Germany joined later in 2014.

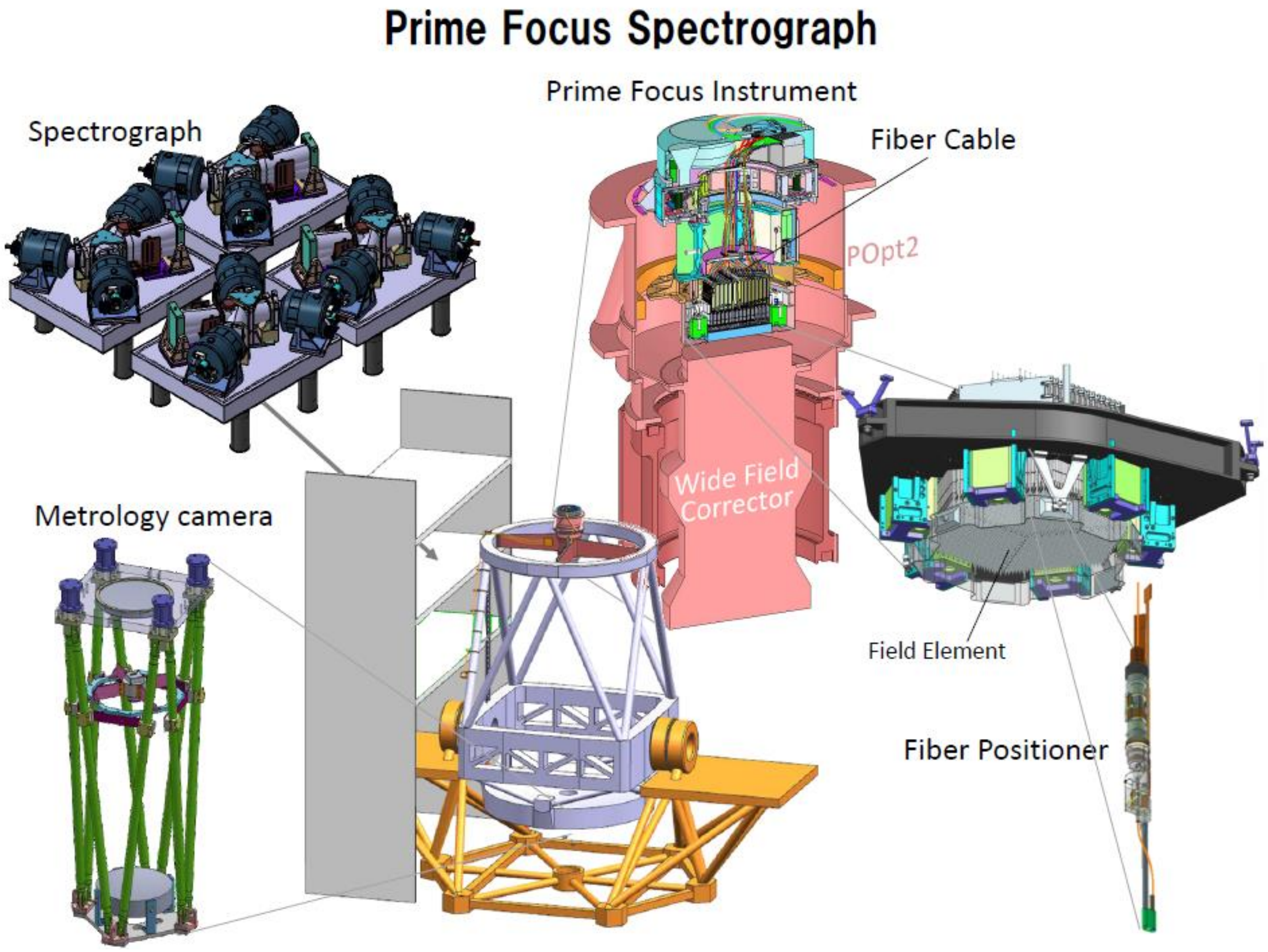

Figure 1. Schematic overview of the Prime Focus Spectrograph (PFS) instrument (not to scale). Subaru's primary mirror focuses light through the wide field corrector (WFC) and a flat glass field element, then onto fibers mounted in an array of fiber positioners. These positioners enable the fiber to be placed on the image of an astronomical object. The fiber positioners as well as other components are mounted into a telescope interface called POpt2. These elements near the telescope focal plane collectively comprise the prime focus instrument (PFI). The fibers are gathered into a fiber cable to route the light to four spectrographs. Each spectrograph also contains backillumination to enable the fiber positions to be measured by the metrology camera.

Table 1. Basic characteristics for Prime Focus Spectrograph (PFS).

| | |
|---|---|
| **Field element** | |
| Shape | Plane; thickness 54 mm |
| **Fiber** | |
| Number | 2394 science fibers |
| Diameter (Core/Cladding/Buffer) | Cable C: 127 µm / 165 µm / 189 µm |
| | Core size corresponds to 1”.12 at field center and 1”.02 at corner when a microlens attached |
| | Cable B: 128 µm / 169 µm / 189 µm |
| | Cable A: 129 µm / 168 µm / 189 µm |
| Length | Cable C: ~5 m; Cable B: ~50 m; Cable A: ~2m |
| Bundling | Cable C: 21 segmented tubes protected by 1 expandable tube |
| | Cable B: 21 segmented tubes bundled and fixed by an adhesive tape around 1 supporting tube |
| | Cable A: 600 science fibers arranged linearly in each of 4 slits |
| Connectors | Two positions: on telescope spider & on spectrograph benches |
| **Microlens** | |
| Shape | Plano-concave (spherical); thickness 3 mm; glued to fiber input edge |
| F-ratio transformation | F/2.2 to F/2.8 |
| **Fiber positioner** | |
| Positioning mechanism | Two stages of piezo-electric rotary squiggle motors |
| Positioner distribution | Hexagonal pattern |
| Distance from neighboring positioners | 8.0 mm |
| Patrol region | 9.5 mm diameter circle for each fiber |
| **Field shape & size** | |
| Field shape | Hexagon |
| Field size/area | Diagonal line length & area |
| | i) Hexagon defined by fiber positioner centers (i.e., twice of distance from field center to farthest fiber positioner centers): 448.00 mm = 1.366 deg on sky. 1.21 $deg^2$ area. |
| | ii) Hexagonal patrol region within which any astronomical target can be accessed at least with one fiber: 453.92 mm = 1.383 deg on sky. 1.24 $deg^2$ area. |
| | When obscured areas of calibration dots mentioned in subsection 2.3 are considered, the above areas should be reduced by 4%. |
| **Spectrograph** | |
| Number | 4 spectrograph modules, |
| | each with a slit of 600 or 597 science fibers & 3-color arms: located on fourth floor infrared side |
| Slit length | ~140 mm, with center-to-center fiber spacing of 213.93 µm |
| F ratios | All-Schmidt type: collimator F/2.5 & camera F/1.09 |
| Grating | VPH; diameter 280 mm |
| Wavelength region | 380-1260 nm (blue: 380-650 nm; red: 630-970 nm; NIR 940-1260 nm) |
| Spectral resolution | ~2.7 A (~1.6 A for red-arm medium resolution mode in 710-885 nm) |
| **Dewar & Detector** | |
| Dewar window | Camera Schmidt corrector |
| Pixel size | 15 µm |
| Detector | A pair of 2K x 4K fully depleted CCDs for each of blue & red arms; |
| | 4K x 4K HgCdTe (1.7 µm cutoff) for NIR arm |
| **Metrology camera** | |
| Location | At Cassegrain |
| Magnification | 0.0366 |
| Camera aperture size | 380 mm |
| Detector | 50M 3.2µm-pixel CMOS sensor |

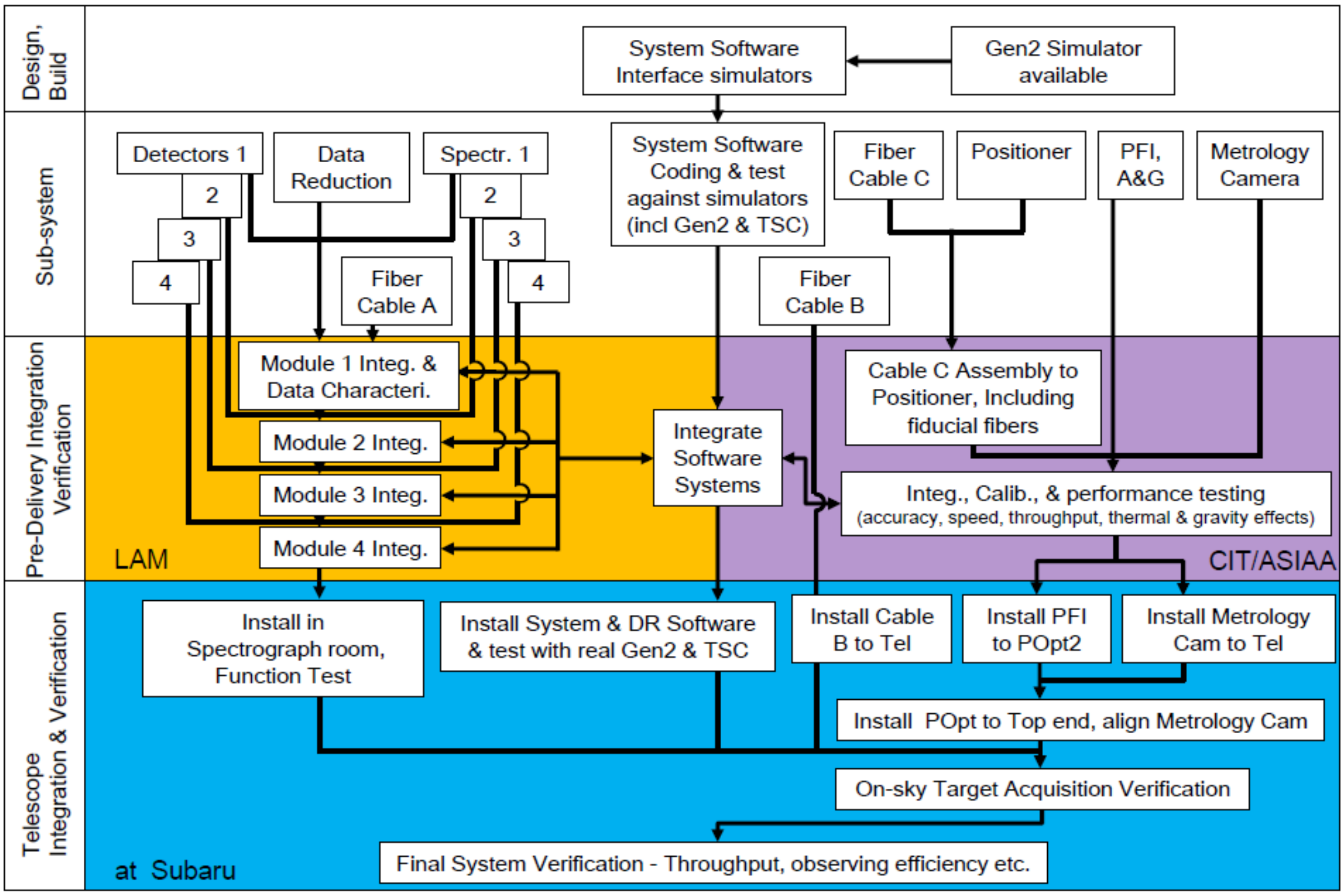

Figure 2. PFS integration and test flow. This flow highlights the modular nature of PFS; the Spectrograph and PFI components can be built and tested independently. The fiber system connects these components.

## 2. DESIGN AND PRODUCTION

Although the detailed design of the instrument has been accomplished through a number of iterations, the most important initial optical parameters were based directly from the scientific case and basic technical constraints. The fiber core diameter of about 100 μm was determined by the image scale at the prime focus provided through WFC and the typical target galaxy size. The numerical aperture, ~0.22 of the fiber, was constrained by the Fratio of the beam provided by WFC. The spectrograph F-ratio of F/1.1 was constrained by the detector pixel size of 15 μm for both the optical and near-infrared (NIR) arms. Since a large demagnification factor is required to avoid oversampling the fiber core images, the ~2400 fibers are separated into four distinct spectrograph modules in order to ensure the required imaging capability of each camera. These first-order constraints were complicated by both F-ratio degradation of a fiber and WFC effects such as nontelecentricity and vignetting; together, these effects require the addition of a microlens on each fiber in the telescope focal plane to transform the input F-ratio from 2.2 to 2.8.

## 2.1 Microlens

A glass-molded microlens, produced by Panasonic Industrial Devices Nitto, transforms the F-ratio F⁄2.2 of the Subaru telescope plus WFC at prime focus into a slower F-ratio of F⁄2.8. This transformation eases the spectrograph design and also ensures an efficient light acceptance cone for each fiber. Glass molding is an attractive fabrication technique because it provides a wider selection of refractive indices than plastic molding and better uniformity of mechanical dimensions than polishing. We selected a high refractive index glass K-VC82 ($n_d$ = 1.75550) for producing the 3-mm thickness microlenses with a 4.764-mm curvature-radius concave at the entrance surface (Fig. 3). A thickness of 3 mm was chosen as a tradeoff between manufacturing constraints and performance for extended targets. The measured thickness uniformity of mass-produced microlenses was within the specification of +/-10 μm. The measured curvature radius and outer diameter of mass-produced microlenses were 4.77 +/- 0.03 mm (1 sigma) and 1.486mm +/- 1.2 μm (1 sigma), respectively. The concave surface has been broadband coated (reflectance <1% between 380 nm and 1.26 μm) and overcoated with $MgF_2$. The opposite flat side will be glued to a fiber. The excellent mechanical uniformity of the microlens enables passive alignment between the fiber and the microlens. We use a black zirconia fiber arm for integrating a fiber, a microlens and a fiber positioner motor shaft.

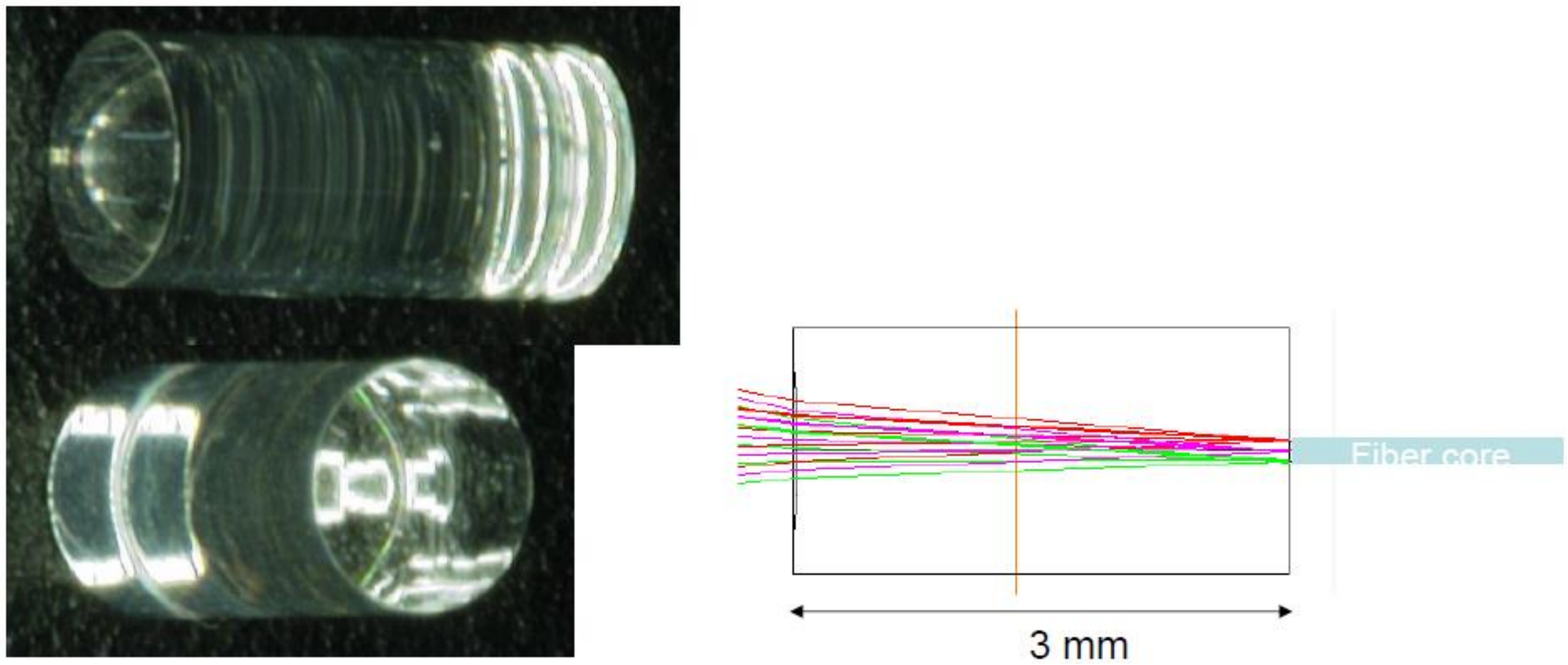

Figure 3. Glass-molded microlens will be attached to the entrance aperture of each fiber for F-ratio transformation.

## 2.2 Fiber system

The fiber system consists of three components termed cables C, B, and A. Cable C is connected to the fiber positioners, capturing the science light in the telescope focal plane. Cable B allows this light to propagate through the telescope structure and building to a special room containing the spectrographs. Cable A is connected to the spectrographs. These cables are linked by two fiber connectors: one on the telescope spider and the other on each spectrograph module bench. This arrangement allows most of the total fiber lengths, cable B, to be permanently installed on the telescope. It permits the focal plane assembly to be removed so that other instruments can be mounted. Finally, it allows easier and safer installation of spectrograph modules and the system separability, as highlighted in Fig. 2. Furthermore, this arrangement simplifies future upgrades and flexibility; for example, a high resolution spectrograph could be realized relatively easily if desired, making use of the existing Cobra fiber positioners and all of the fixed fiber systems.

Each cable is terminated with an assembly based on custom, 32-fiber, MTP connectors produced by US Conec, Ltd. On the telescope spider, cable B is connected to cable C using 22 USCONEC connectors in an assembly called the "tower connector"; four tower connectors are needed to connect all 2394 fibers. The tower connector is engineered to be particularly thin to meet challenging restrictions on size and shape at the prime focus. At the spectrograph end, cable B is connected to cable A using 12 USCONEC connectors in an assembly of eight so-called "gang connectors."

Fiber selection has been one of the key issues in the design of the fiber system. The fibers must have both excellent transmission across a wide spectral band (380–1260 nm) and low FRD properties. The project has intensively tested and evaluated fibers produced by two companies: a Japanese company Fujikura and a US company Polymicro. Two methods were used for the FRD measurements (Fig. 4). In the first method, as shown in Fig. 4(a), monochromatic light (filtered for 550 nm or He-Ne laser 633 nm) was input at the nominal PFS aperture, F∕2.8. In the second method, shown in Fig. 4(b), a collimated beam was input and scanned with the incident angle. Measurements showed that both fibers met system requirements for FRD and transmission. The Polymicro fibers were found to have a slightly better FRD performance. A Polymicro 50-m fiber had about 1% loss from FRD at F∕2.5, while a similar Fujikura 50-m fiber had about 3% loss. Fujikura fibers provide slightly better transmission: the gain depends on wavelength but is 2%–10% higher for a 50-m length.

Based on the above results, we selected Polymicro fibers for cables C and A, and Fujikura fibers for cable B. Cable B is by far the longest run, so this ensures a high overall transmission. Cables A and C have tighter geometrical constraints and are, therefore, more susceptible to FRD losses making the Polymicro fiber the natural choice.

In order to minimize geometrical losses at the connectors, the outer polyimide-buffer diameter of the fibers was specified as 189 +/- 3 μm for Fujikura and 190 + 3 ∕ - 5 μm for Polymicro. This ensures better center alignments between fibers at each USCONEC connector, whose hole diameter specification is 195 +/- 1 μm. To further reduce light loss, we specified core (and clad) diameters as 127 μm +/- 3 μm (165 +/- 3 μm), 128 μm +/- 5 μm (169 +/- 2 μm), and 129 +/- 3 μm (168 +/- 3 μm), respectively, for cables C, B, and A. This flow from a smaller core diameter to a larger core diameter (127 < 128 < 129) will minimize geometrical losses in a statistical way, even with slight misalignments of core centers between upstream and downstream fibers, as well as with the possibility of a few μm offset between the core center and buffer center. This intentional slight change of core diameters at the connectors will slightly affect the F-ratio, but the effect of geometrical area matching at connectors will be more significant than that of the slight F-ratio change. Both of Fujikura and Polymicro fiber productions have already been completed.

To optimize uniformity of fiber image qualities, the mapping of fibers between the fiber positioner and the locations in the fiber slits has been carefully optimized. The focal plane has its best image quality near the center of the field and has a significant loss of telecentricity near the edge. Similarly, the spectrograph has better image quality near the center of the FOV. To balance these two effects, fibers from the edge of the telescopic FOVare routed to the inner location of a spectrograph slit, and fibers closer to the center of the telescopic FOV are routed to the outer location of the spectrograph slit. Such a fiber-routing pattern has been successfully designed, maintaining the modularities of fiber positioners and of spectrographs.

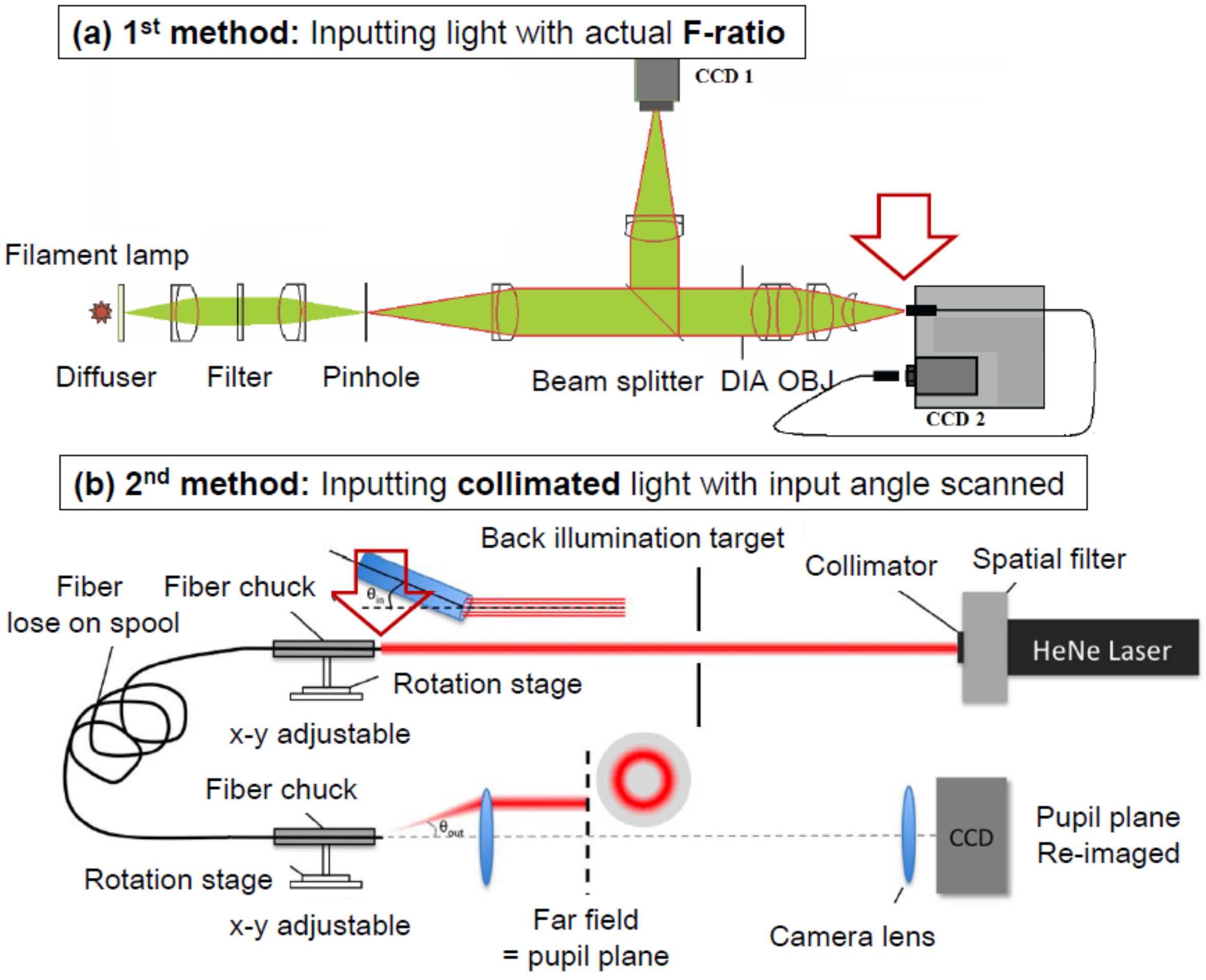

Figure 4. Two methods used for focal ratio degradation (FRD) measurements. (a) The F-ratio of 2.8 is input at the fiber entrance indicated by a red arrow and the total effect of FRD is measured. The combination of diaphragm (DIA) and a camera lens (OBJ) produces the required F ratio. A camera indicated as CCD2 measures the angular distribution of the output light from a fiber, while another camera is used for monitoring and alignment purposes. (b) Collimated light is input (shown by a red arrow) and its incident angle is scanned by using a rotating stage, so that the effects on the specific incident angles are measured. The angular distribution of the output light from the fiber is relayed onto a charge-coupled device (CCD) camera through a re-imaging lens. Both methods are valid techniques and were used to improve confidence in the final selection of fiber manufacturer. The upper figure was taken from de Oliveira et al.'s[20] paper. The measurements made using the latter method were carried out by Murray, Dunlop, and Allington Smith at Durham University CfAI, and the lower figure they kindly provided was originally produced by Ulrike Lemke, who is currently at the Institut für Astrophysik, Göttingen.

## 2.3 Fiber positioner and Prime focus instrument

We use fiber positioner units developed by JPL in partnership with New Scale Technologies. They consist of two-staged piezo-electric rotary squiggle motors arranged so as to place an individual fiber onto a target object within a particular patrol area on the focal plane. Intensive life time tests on such designmodified positioners are being carried out in 300 k cycles of clockwise plus counterclockwise rotations including hard stop events and some realistic movements at a temperature of −5°C, i.e., the most severe environment expected on Mauna Kea.

Since open-loop positioning accuracy is insufficient, we carry out iterations based on the fiber position measured by a metrology camera, which is described in Sec. 2.5. Figure 5 shows an example of how the fiber position converges onto the target position for 100 randomly selected target points within the patrol area of 9.5-mm diameter. Intensive tests using 29 engineering model fiber positioners have been completed and we are now in the phase of mass production (Fig. 6).

The fiber positioners are built into a module containing 57 positioners each with a microlensed fiber and the Cobra drive electronics. The total positioner system consists of 42 of these modules mounted onto the fiber-positioner optical bench, which is attached to the PFI structure. As shown in Fig. 1, the PFI is installed in Subaru's housing structure, called prime focus unit POpt2, whose rotator provides PFI with the rotating part and the nonrotating part with respect to the telescope structure. While the nonrotating part is fixed to the telescope, rotation enables tracking target objects in the sky. A structure called the positioner frame connects the POpt2 rotator and the fiber-positioner optical bench on which fiber positioners as well as six acquisition and guide (A&G) cameras are mounted. For each of A&G cameras, we use FLI ML 4720, a1 K×1 Kframe transfer camera with 13 μm pixels. Its coverage of 5.5 $\text{arcmin}^2$ on the sky is wide enough for providing at least one star for the A&G in a reasonable exposure time even in the lowest star-density area. Ninety-six fiducial fibers are mounted on this optical bench and provide the metrology camera with reference positions when they are artificially back-illuminated. For this purpose, the fiducial fiber illuminator is also located in the PFI rotating part. The cable wrapper ensures cables of fibers/electronics/ coolant behave in a well-organized way when the rotator moves.

A glass plate, called the field element, is attached to the fiberpositioner optical bench by three field element mounts. The thickness of the field element, 54 mm, roughly corresponds to the total thickness, 52 mm, of a HSC dewar window and a filter that are removed during the instrument exchange from the HSC to the PFS. The field element has an array of thin chrome-deposited dots at the surface closest to the image plane to enable better calibration. It is possible to obtain each pure fiber image or each pure spectrum on the spectrograph detectors by hiding some of fibers from calibration sources, such as a flat-field lamp and a wavelength calibration lamp. The dot size has been determined to be 1.5 mm, considering not only the geometrical obscuration at the distance of 500 μm between the field element and tips of microlenses but also the open-loop accuracy of last fiber positioning. The production of the field element with dots has been completed.

Calibration lamps are mounted on top of the PFI structure. To ensure the uniformity of lamp illumination, we will have a single flat-field lamp as well as wavelength calibration lamps (one for blue and another for red/near-infrared). These lamps illuminate a Lambertian flat-field screen attached at the bottom of the dome shutter. The desired irradiance pattern is obtained by using strongly aspheric plastic lenses in the illumination optics. Figure 7 summarizes all the above functions contained in the PFI.

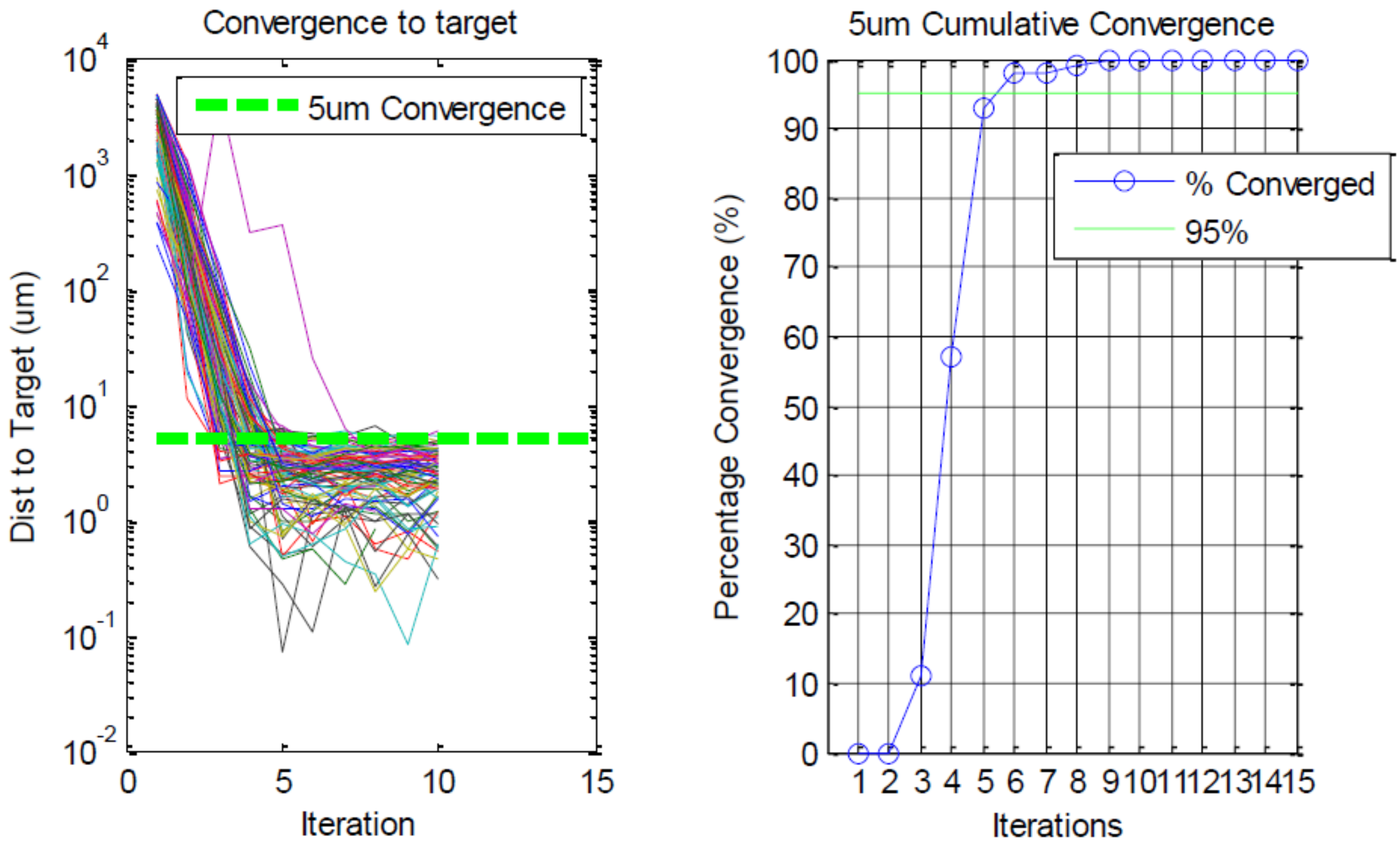

Figure 5. Convergence to target for the best (among five at the time) engineering model (EM) fiber positioner. The results shown are for 100 randomly selected target points within the 9.5-mm diameter patrol area. (a) shows how the actual fiber position is converged with iteration onto the target position for each of these 100 target positions (cases), (b) shows the fraction of cases where the fiber positions are converged within 5 μm with respect to the target positions.

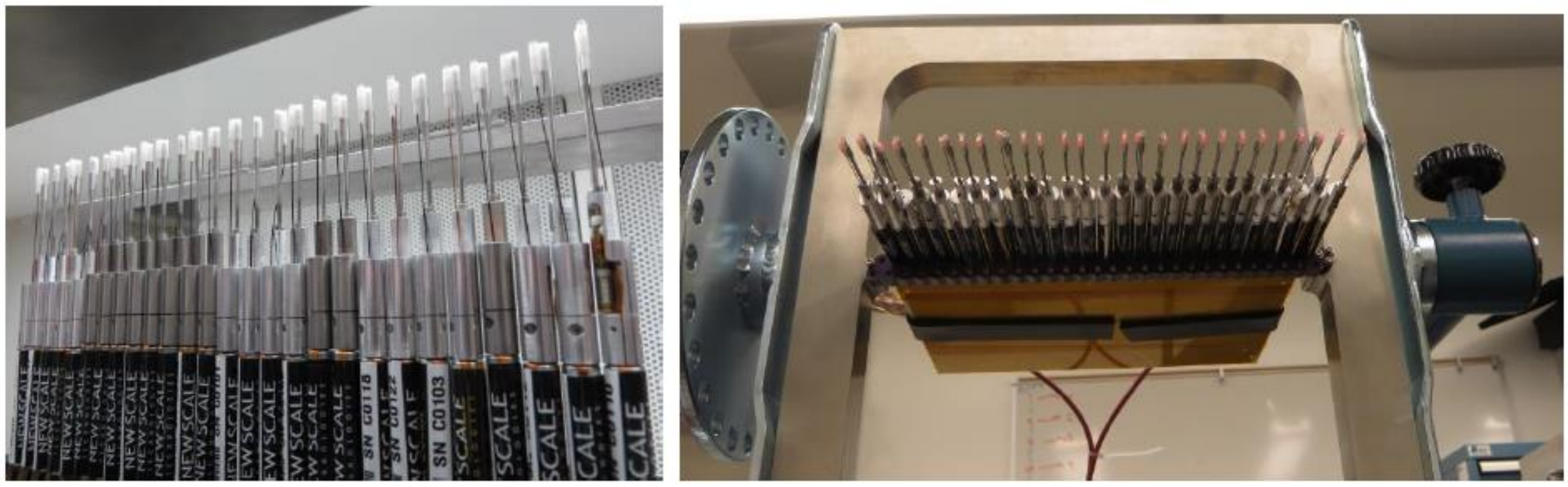

Figure 6. EM fiber positioners arranged in a module. The final module will contain 57 positioners in two adjacent rows. Many aspects of operating a number of positioners have simultaneously been tested, including collisional avoidance between fiber arms and a crosstalk check within an electronic board and/or adjacent electronic boards. For scientific operations, the PFI control software will provide target and current positions of each fiber positioner to the Cobra movement planning software, which generates parallel sets of commands to move individual positioners to their targets after computing a safe path for each positioner. The baseline approach to avoid collisions between positioners is to run each positioner radially when transiting common patrol areas.

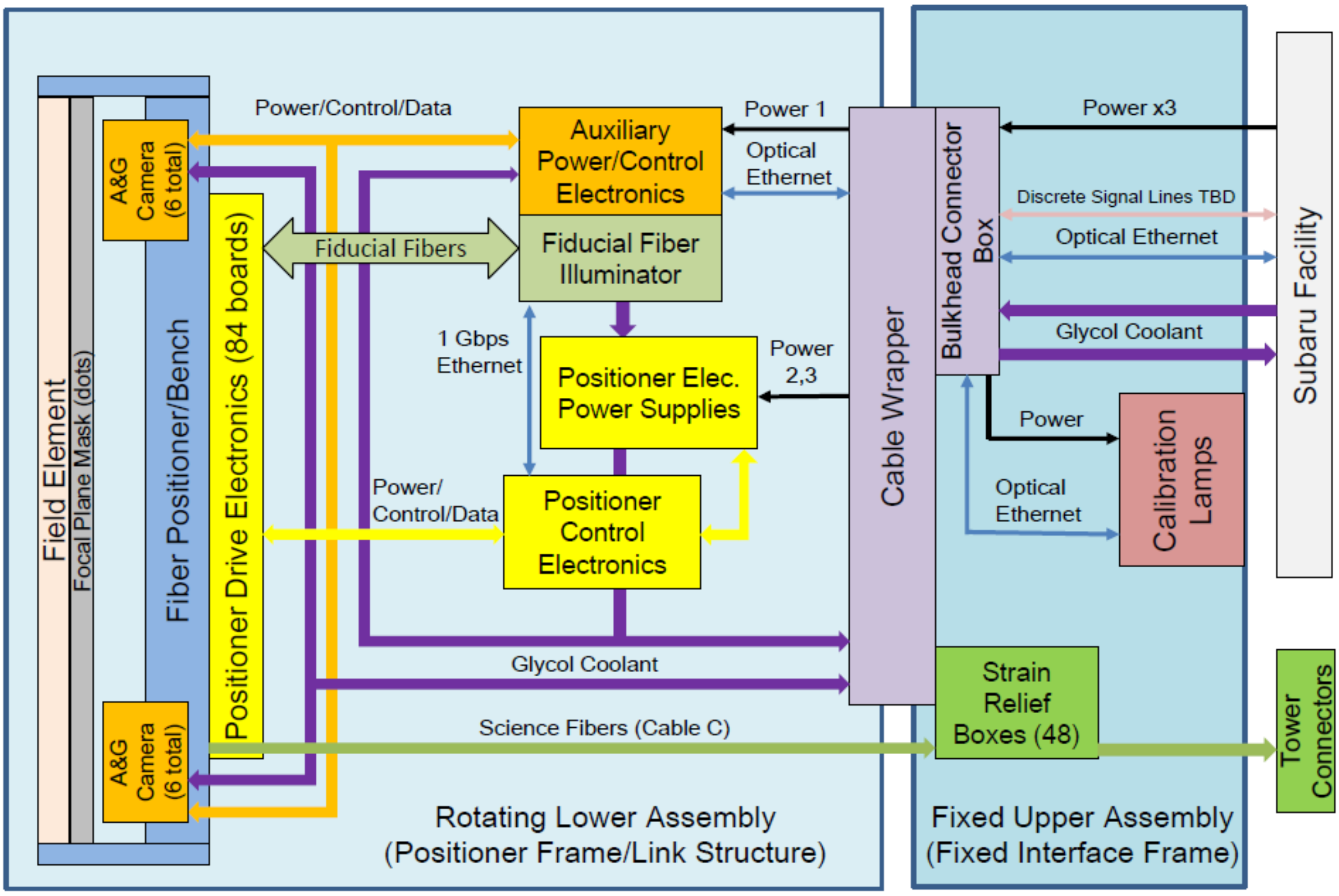

Figure 7. Functional block diagram for PFI. The upper assembly is fixed with respect to the telescope structure, while the rotating lower assembly through the cable wrapper provides tracking target objects in the sky. The calibration lamps are in the fixed assembly, while the field element with dots, fiber positioners, fiducial fibers, and A&G cameras are in the rotating assembly.

## 2.4 Spectrograph

The optical design of a double Schmidt-type spectrograph has been completed, with the 380 nm − 1.26 μm wavelength range divided into three arms by two dichroic mirrors. Figure 8 shows one of the four such identical spectrograph modules. The slit structure at the entrance of each spectrograph module uses holes in front- and rear- thin electroformed nickel masks to define the fiber–fiber distances as well as their alignment. The arrangement is complicated by the requirement to allow for a nontelecentric slit plane. This requirement is driven by the 280-mm diameter limit on the grating fabrication. After developing prototype slit assemblies and measuring thermal effects on their stabilities, we are now completing the first final slit assembly. The production of the spectrograph optical elements is also in progress. All silica glass blanks are ready and being polished by Winlight System. Figure 9 shows an example of the precision optics required for the spectrographs. This 305 mm × 600 mm × 60 mm thick optic meets our stringent requirements: radius of curvature = 1389 mm, a measured surface error <130 nm peak-to-valley (P-V) for each beam footprint, and root mean square (RMS) surface roughness <1.3 nm. Figure 10(a) shows one of the three prototype volume phase holographic (VPH) gratings produced by Kaiser. Although these are prototypes, they have qualities close to our specification: e.g., the RMS wavefront error for the blue VPH grating prototype was 250 nm for the 280-mm clear aperture, with a trefoil pattern as the dominant error source.We have recently identified the main cause of this trefoil pattern as a slight misalignment of optics for VPH production. This misalignment is now fixed and we are starting mass production of VPH gratings. Figure 10(b) shows an Ohara high-refractive-index prism blank made of S-LAH53 for a medium resolution grism in the red channel. Such a blank with high refractive index of $n_d = 1.80610$ and with a large size of 320-mm diameter has been successfully produced, realizing the homogeneity of $4 \times 10^{-6}$ P-V for the whole clear aperture when the spherical component is removed. This blank has been cut into two prisms, each of which will be glued to either side of a medium resolution VPH grating to form a medium resolution grism. The exchange between low resolution mode and medium resolution mode can be done remotely by using a slidetype exchanger of red gratings/grisms.

For the red channel in each spectrograph module, we use a pair of 2 K×4 K Hamamatsu fully depleted charge-coupled devices (CCDs) with a pixel size of 15 μm × 15 μm. This is exactly the same CCD used for the HSC. To simplify the control electronics system, mechanical interfaces, as well as operation on optical channels, a pair of 15 μm × 15 μm-pixel 2 K×4 K Hamamatsu fully depleted CCDs is also used for the blue channel but with a thinner fully depleted region and a blue optimized coating. The 100-μm depletion layer provides a smaller charge diffusion of about 4 μm (1 sigma) compared with 7.5 μm for a standard thickness of 200 μm. The blue optimized coating provides a detector quantum efficiency of 57%, 74%, and 91% at 380, 400, and 500 nm, respectively, a significantly improvement over a standard coating. All the CCDs have been successfully produced. In collaboration with NAOJ Advanced Technology Center (ATC), AlN pin bases were prepared for individual optical CCDs by placing Ti pins with a positional accuracy sigma = 1.4 μm in each of the x, y directions and a pin-diameter accuracy of 0.4 μm. These pins allow accurate alignment of a pair of CCDs. For the NIR channel in each spectrograph module, a 4 K×4 K Teledyne H4RG-15 mercury-cadmium telluride device is used with the same pixel size of 15 μm × 15 μm to the optical CCD case. The special cutoff wavelength of 1.7 μm, as short as is feasible, suppresses the thermal background.

The expected location for installing the spectrograph modules is the fourth floor of the infrared side (IR4) in the Subaru enclosure. The four spectrograph modules will lie within a single light-weighted spectrograph clean room, with thermal insulator walls to improve both temperature control and cleanliness. The temperature inside of the spectrograph clean room will be kept within +/-1°C. This thermal control stabilizes image quality and limits drift of image position to less than 0.2 pixel over 1 h. The optical bench for each spectrograph module is made of carbon fiber reinforced plastic (CFRP), which has a low coefficient of thermal expansion, further stabilizing the spectrograph optics configuration against residual temperature variation. The selection of CFRP has also been essential for reducing weight, as compared with stainless steel.

The effects of possible vibrations have also been thoroughly investigated. For cooling the dewars and their enclosed detectors, we use Sunpower compact free-piston stirling cryocoolers. These have been selected because of cost and reliability; the mean time between failures (MTBF: the predicted elapsed time between inherent failures of a system during operation) is about 14 years. The vibration amplitude produced by the cryocooler is reduced, e.g., down to −40 dBG RMS level at the 60 Hz fundamental with an active damper. The project, however, has carefully designed the mechanical and optical components so that their resonance frequencies are larger than 85 Hz for all components when attached to the optical bench and are larger than the second harmonic (120 Hz) for internal modes except for the heaviest component cases, such as a grating/grism exchange mechanism.

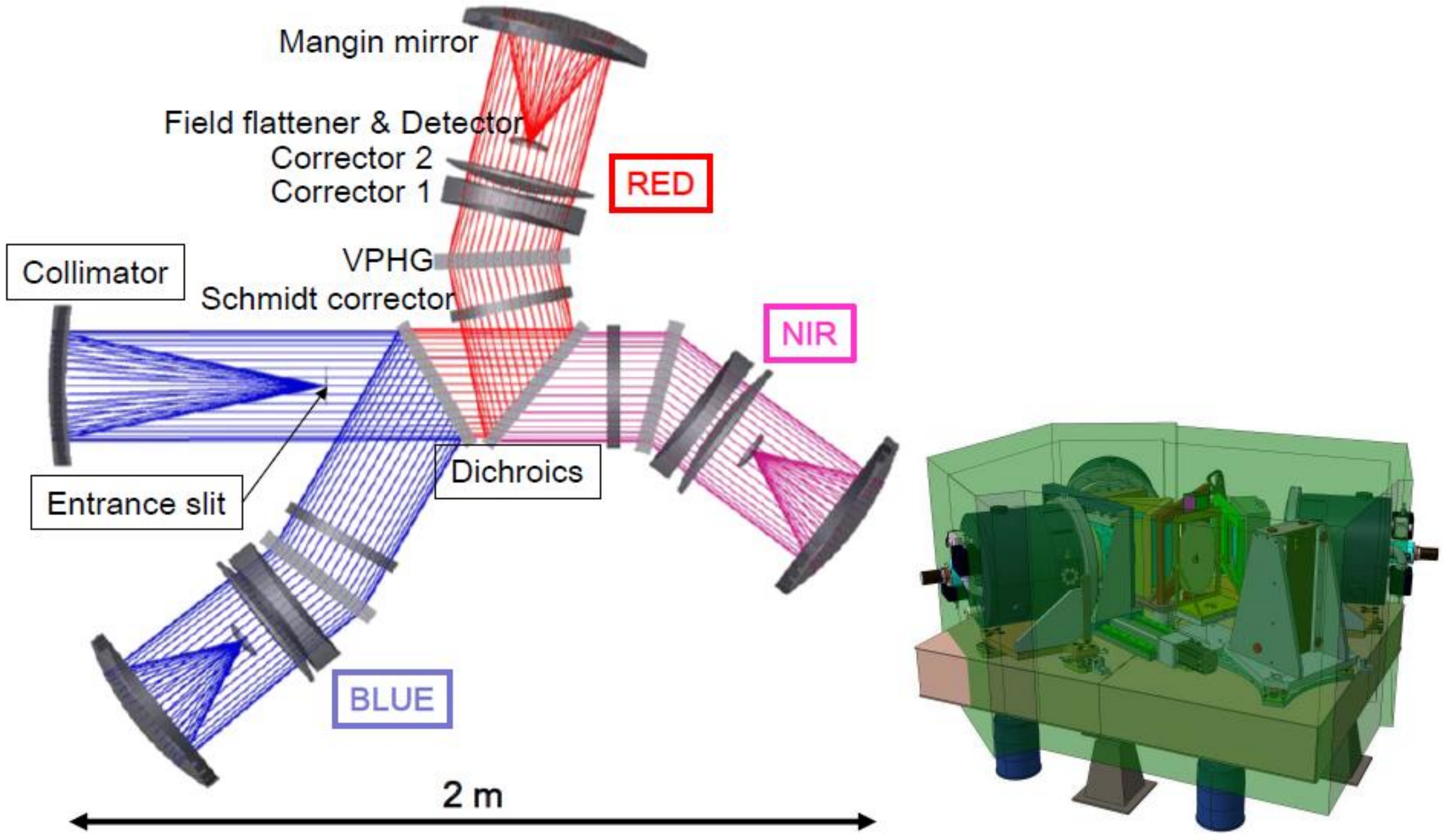

Figure 8. One of the four identical spectrograph modules. Schmidt-type optics are used both on the F/2.5 collimator and the F/1.09 camera. The entrance slit is vertical to the page and the distance from the slit to the collimator is 693.5 mm. The three arms have almost identical optical elements, so only those for the red arm are labeled. Corrector 1 for each color arm works also as the dewar window. The medium resolution mode for the red channel will be realized by using a slide structure (not shown) with a low resolution grating and a medium resolution grism. The scale shown at the bottom left is only for the left drawing.

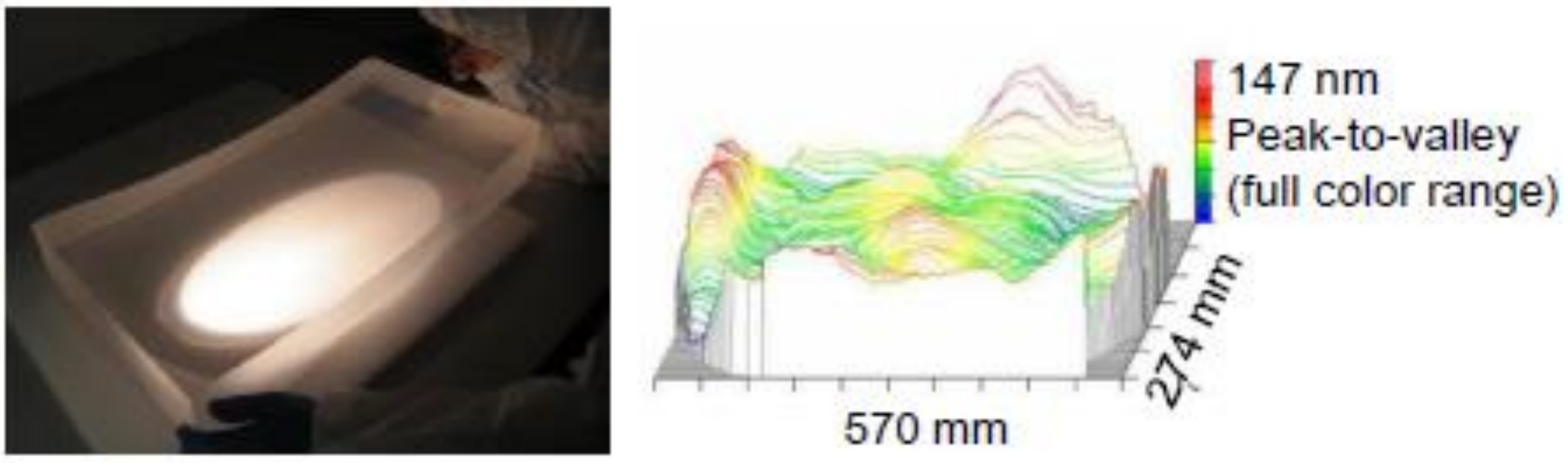

Figure 9. A polished collimator mirror. The size is 305 mm × 600 mm × 60 mmthickness at its center. The measured surface shape error was 147 nm peak-to-valley (P-V) for the whole effective area and was less than 130-nm peak-to-valley per each fiber pupil.

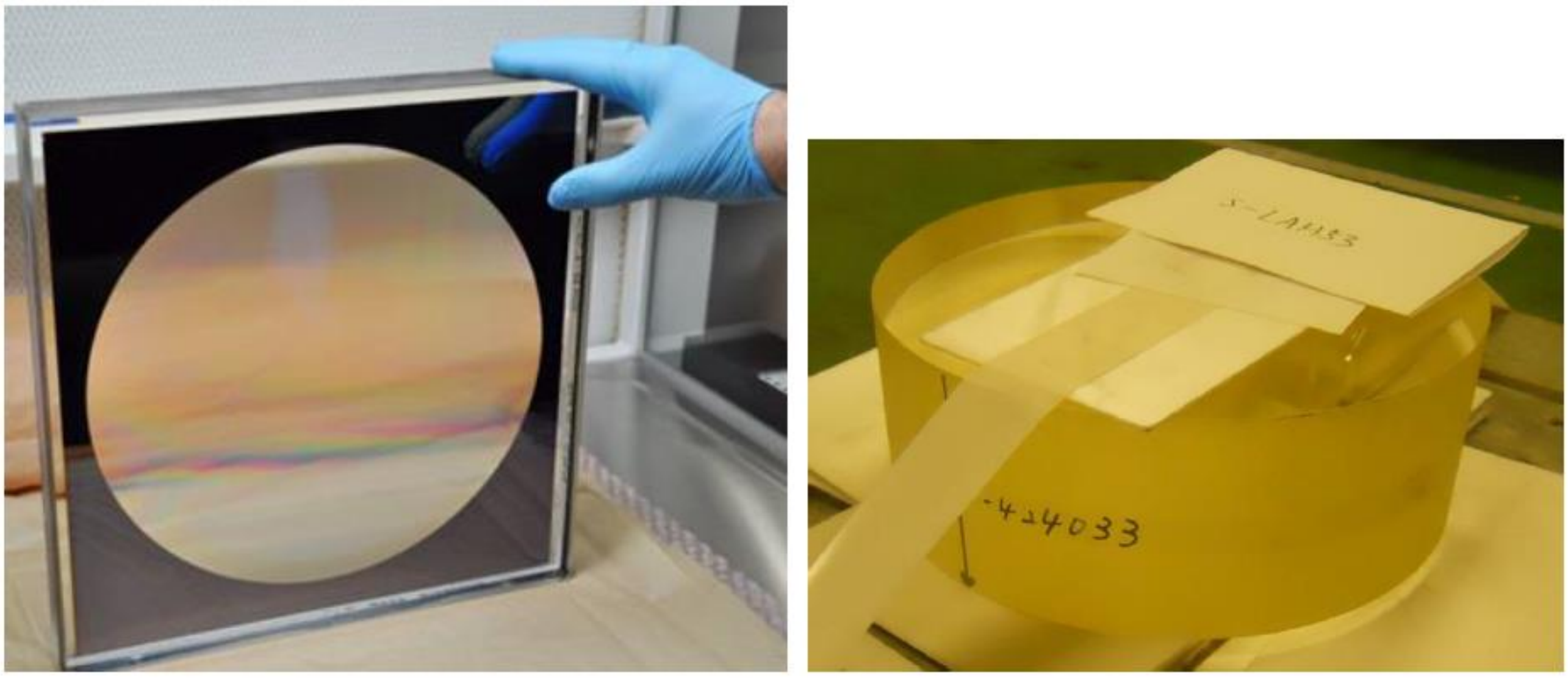

Figure 10. (a) Prototype VPH grating for the blue channel. The outer part of the grating is masked with chrome, which provides us the only real light stop in the optical system. The size of the grating is 340 mm × 340 mm × 40 mm thickness. (b) A blank for a pair of prisms that will be attached to a volume phase holographic (VPH) grating to form a red-channel medium-resolution grism. The diameter of the blank is 320 mm.

## 2.5 Metrology camera

As described in subsection 2.3, iteration is necessary to accurately position each fiber. To facilitate this iteration, the spectrograph is equipped with light emitting diodes that provide back-illumination when the spectrograph shutter is closed. These back-illuminated fibers are imaged by a metrology camera located at Cassegrain. The metrology camera is set in the common-use Cassegrain container,[21,22] which has been developed in collaboration between the Kyoto tridimensional spectrograph II[23,24] team and NAOJ. The container is installed with a robotic instrument exchanger Cassegrain instrument auto eXchanger for the Subaru telescope.[25,26] Given the large distance between the fibers and the metrology camera, this configuration permits all fiber images to be viewed in one exposure. To validate the precision of this measurement, the “dome seeing” was measured at the Subaru telescope by using a similar configuration to back-illuminate Fiber Multi-Object Spectrograph[27] fibers imaging with a commercial-based CCD camera Atik 450 set at the common-use Cassegrain container. To average out short-timescale variations in the perceived fiber positions due to “dome seeing,” an exposure time of around 1 s is optimal. Exposure times of 0.5 and 1.0 s lead to RMS image position variations of 2.5 μm and 2 μm, respectively. These variations are acceptably small compared to the final fiber positioning accuracy requirements of ~10 μm.

The diameter of the metrology camera optics was chosen to minimize the effects of small surface deformations within the WFC optics. The optical path from the fiber plus microlens to the metrology camera is much narrower than the path from the telescope primary mirror to fiber plus microlens since the metrology camera aperture is much smaller than the primary mirror aperture (Fig. 11). This difference does not matter much if the WFC lens surface shapes are perfect, because the different paths share the same chief ray. The WFC lenses, however, have surface shape errors including high spatial frequencies around 6 mm with the P–V amplitude of 10–30 nm; these small errors are inherent to the manufacturing process that is required for the aspheric surfaces in the WFC. Such errors are allowed for the HSC imaging since the amplitude is much less than the wavelengths used. However, because the PFS metrology camera uses only a tiny portion of the WFC lenses for each fiber, the local slopes of the lenses produce significant deformation of coordinates of fibers in the focal plane when they are measured in the metrology camera coordinates. To suppress this effect down to the level of standard deviation of 1.7 μm, an aperture diameter of 380 mm is required for the metrology camera; this diameter corresponds to the unobscured maximum aperture at Cassegrain. This large entrance pupil diameter would cause diffraction-limited images of the fibers to be undersampled, preventing accurate centroiding. To prevent this condition, the metrology camera optics were designed with a lot of spherical aberration, ensuring the point spread functions are uniform across the field and are several pixels in diameter.

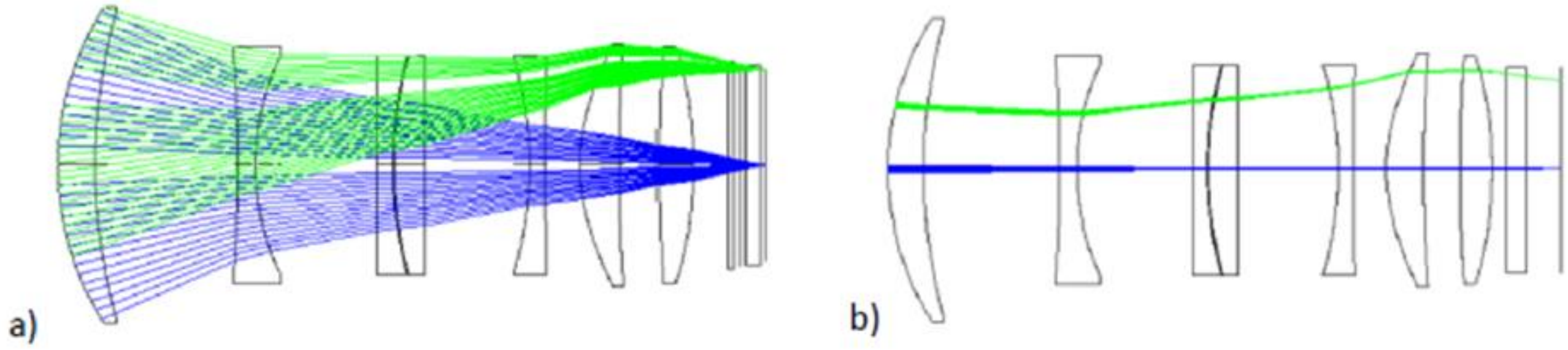

Figure 11. (a) Schematic diagram for the light path (coming from the left side of this figure) of a target object through the Subaru telescope. (b) Light path from the artificially backilluminated fibers (at the right side of this figure) to the metrology camera located at Cassegrain. The metrology camera uses a tiny portion of WFC lenses.

### 2.6 Software and execution processes of observations

For science operations, the observation preparation software parses large surveys into a series of individual observations, matches fibers to designated targets, selects appropriate guide stars, predicts observing times, and provides sequence files in a Subaru format. The PFS observation control computer (OBCP) controls the entire instrument under the Subaru observation management software Gen2. Execution of an observation begins with moving the telescope to the desired field center. While the telescope is slewing, the fiber positioning system executes commands from the OBCP to simultaneously move all fibers to their required positions with the real positions verified by the metrology system. With several iterations, the position error of each science fiber can be effectively minimized. In parallel with this process, the A&G system refines the pointing of the telescope and angle of the instrument rotator. The hexapod is commanded to move to the appropriate position to maintain the WFC alignment in the presence of flexure in the POpt2 structure. Once positioning is complete, autoguiding commences. The onsite data reduction system produces quick-look information on the completed observation so that data quality and survey progress can be monitored. The collected raw data will be transferred to and archived in the Subaru telescope archive system data archive system, which then provides data retrieval for further scientific data reduction and analysis.

## 3. SUMMARY

Following the successful project conceptual design review and preliminary design review and further subsystem critical design reviews, PFS has now entered the construction phase. Mass production of the microlenses and fibers is complete. These two components are now being integrated, which includes gluing them in the fiber arm structure. Later, they will be integrated with each fiber positioner, shortly to enter the mass production phase, by gluing the fiber arm to the positioner motor shaft. The positioner modules will be integrated onto the positioner optical bench and then into the PFI structure. This integrated PFI system will be tested together with the metrology camera. For the spectrograph module side, all optical elements, including gratings, are in production. These optical components will be integrated with cryostats and dewars on four spectrograph optical benches. After acceptance tests of each of the PFI and spectrograph components, they will be transported to Hawaii, where the whole system finally will be tested all together: the PFI system in the POpt2 at the telescope top, spectrograph modules in a single clean room on the fourth floor IR side, and the metrology camera in the Cassegrain container. The technical first light is planned in early 2018.

## ACKNOWLEDGEMENTS

We gratefully acknowledge support from the Funding Program for World-Leading Innovative R&D on Science and Technology (FIRST) program “Subaru Measurements of Images and Redshifts (SuMIRe)”, CSTP, Japan, and Fundação de Amparo a Pesquisa do Estado de São Paulo (FAPESP), Brasil. We appreciate staff members at Subaru Telescope for continuously supporting our activities. We thank NAOJ ATC staff members particularly Tetsuo Nishino, Norio Okada, and Yukiko Kamata for preparing AlN pin bases, and Durham University staff members for their consultancy to IPMU on fiber system. We also acknowledge the WFMOS-B team whose accumulated efforts of many years have inspired us. This paper has been revised and updated, for the purpose of journal publication, from the PFS overview paper[28] presented in an SPIE conference.

## REFERENCES

1. Sugai, H. et al., “Prime Focus Spectrograph – Subaru’s future –,” Proc. SPIE 8446, 84460Y, 13pp. (2012).
2. Miyazaki, S., “Hyper suprime-cam for weak gravitational lensing survey (Presentation Video)”, Proc. SPIE 9143, Space Telescopes and Instrumentation 2014: Optical, Infrared, and Millimeter Wave, 91431Z (July 14, 2014)
3. Takada, M. et al., “Extragalactic science, cosmology, and Galactic archaeology with the Subaru Prime Focus Spectrograph,” Publications of the Astronomical Society of Japan 66(1), R1(1-51) (2014).
4. Takato, N. et al., “Design and performance of a F/#-conversion microlens for Prime Focus Spectrograph at Subaru Telescope,” Proc. SPIE 9147, 914765, 7pp. (2014).
5. de Oliveira, A. C. et al., “Fiber Optical Cable and Connector System (FOCCoS) for PFS/ Subaru,” Proc. SPIE 9151, 91514G, 11pp. (2014).
6. dos Santos, J. B. et al., “Studying focal ratio degradation of optical fibers with a core size of 128 microns for FOCCoS/ PFS/ Subaru,” Proc. SPIE 9151, 915150, 6pp. (2014).

7. de Oliveira, A. C. et al., “Slit device for FOCCoS – PFS – Subaru,” Proc. SPIE 9151, 91514D, 13pp. (2014).
8. de Oliveira, A. C. et al., “Polish device for FOCCoS/PFS slit system,” Proc. SPIE 9151, 915145, 7pp. (2014).
9. de Oliveira, A. C. et al., “Multi-fibers connectors systems for FOCCoS-PFS-Subaru,” Proc. SPIE 9151, 915160, 9pp. (2014).
10. Fisher, C. D. et al., “Developing Engineering Model Cobra fiber positioners for the Subaru Telescope’s Prime Focus Spectrometer,” Proc. SPIE 9151, 91511Y, 13pp. (2014).
11. Wang, S. -Y. et al., “Prime Focus Instrument of Prime Focus Spectrograph for Subaru Telescope,” Proc. SPIE 9147, 91475Q, 9pp. (2014).
12. Vivès, S. et al., “Current status of the Spectrograph System for the SuMIRe/PFS,” Proc. SPIE 9147, 914762, 9pp. (2014).
13. Pascal, S. et al., “Optical design of the SuMiRe/PFS Spectrograph,” Proc. SPIE 9147, 914747, 8pp. (2014).
14. Madec, F. et al., “Integration and test activities for the SUMIRE Prime Focus Spectrograph at LAM : first results,” Proc. SPIE 9147, 91475Z, 9pp. (2014).
15. Hart, M. et al., “Focal Plane Alignment and Detector Characterization for the Subaru Prime Focus Spectrograph,” Proc. SPIE 9154, 91540V, 16pp. (2014).
16. Hope, S. et al., “CCD readout electronics for the Subaru Prime Focus Spectrograph,” Proc. SPIE 9154, 91542G, 14pp. (2014).
17. Barkhouser, R. H. et al., “Volume phase holographic gratings for the Subaru Prime Focus Spectrograph: performance measurements of the prototype grating set,” Proc. SPIE 9147, 91475X, 18pp. (2014).
18. Gunn, J. E. et al., “The near infrared camera for the Subaru Prime Focus Spectrograph,” Proc. SPIE 9147, 91472V, 26pp. (2014).
19. Wang, S. -Y. et al., “Metrology camera system of Prime Focus Spectrograph for Subaru telescope,” Proc. SPIE 9147, 91475S, 10pp. (2014).
20. de Oliveira, A. C. et al., “FOCCoS for Subaru PFS” Proc. SPIE 8446, 84464R, 1-14 (2012).
21. Sugai, H., Hattori, T., Kawai, A., Ozaki, S., Kosugi, G., Ohtani, H., Hayashi, T., Ishigaki, T., Ishii, M., Sasaki, M., Shimono, A., Okita, Y., Sudo, J. and Takeyama, N., “Test Observations of the the Kyoto Tridimensional Spectrograph II at University of Hawaii 88-inch and Subaru Telescopes,” Proc. SPIE 5492, 651-660 (2004).
22. Sugai, H., Ohtani, H., Ozaki, S., Hattori, T., Ishii, M., Ishigaki, T., Hayashi, T., Sasaki, M. and Takeyama, N., “The Kyoto Tridimensional Spectrograph II: Progress,” Proc. SPIE 4008, 558-569 (2000).
23. Sugai, H., Hattori, T., Kawai, A., Ozaki, S., Hayashi, T., Ishigaki, T., Ohtani, H., Shimono, A., Okita, Y., Matubayashi, K., Kosugi, G., Sasaki, M. and Takeyama, N., “The Kyoto Tridimensional Spectrograph II on Subaru and the University of Hawaii 88 in Telescopes,” Publications of Astronomical Society of the Pacific 122, 103-118 (2010).
24. Sugai, H., Ohtani, H., Ishigaki, T., Hayashi, T., Ozaki, S., Hattori, T., Ishii, M., Sasaki, M. and Takeyama, N., “The Kyoto Tridimensional Spectrograph II,” Proc. SPIE 3355, 665-674 (1998).
25. Usuda, T. et al., “CIAX: Cassegrain instrument auto exchanger for the Subaru telescope,” Proc. SPIE 4009, 141-150 (2000).

26. Omata, K. et al., “Control of the Subaru telescope instrument exchanger system,” Proc. SPIE 4009, 374-385 (2000).
27. Kimura M., Maihara T., Iwamuro F., Akiyama M., Tamura N., Dalton G. B., Takato N., Tait P., Ohta K., Eto S., Mochida D., Elms B., Kawate K., Kurakami T., Moritani Y., Noumaru J., Ohshima N., Sumiyoshi M., Yabe K., Brzeski J., Farrell T., Frost G., Gillingham P. R., Haynes R., Moore A.. M, Muller R., Smedley S., Smith G., Bonfield D. G., Brooks C. B., Holmes A. R., Curtis Lake E., Lee H., Lewis I. J., Froud T. R., Tosh I. A., Woodhouse G. F., Blackburn C., Content R., Dipper N., Murray G., Sharples R. and Robertson D. J., “Fibre Multi-Object Spectrograph (FMOS) for the Subaru Telescope,” Publications of the Astronomical Society of Japan 62, 1135-1147 (2010).
28. Sugai, H. et al., “Progress with the Prime Focus Spectrograph for the Subaru Telescope: a massively multiplexed optical and near-infrared fiber spectrograph,” Proc. SPIE 9147, 91470T, 14pp. (2014).